\documentclass[runningheads,a4paper]{llncs}

\usepackage{amsmath,amsfonts,graphpap,amscd,mathrsfs,graphicx,lscape}
\usepackage{epsfig,amssymb,amstext,xspace}
\usepackage{algorithm,algorithmic}

\usepackage{color}              % Need the color package
\usepackage{epsfig}

\newtheorem{defn}[theorem]{Definition}
\def\squareforqed{\hbox{\rule{2.5mm}{2.5mm}}}

\def\QED{\ifmmode\squareforqed % in mathmode : print just the square
  \else{\nobreak\hfil   % \hfil to end of current line
    \penalty50                 % penalty 50 for breaking the line here
    \hskip1em                  % leave at least 1em before the square
    \null                      % \hbox{}
    \nobreak                   % prohibit line break
    \hfil                      % another \hfil (if a break occurred)
    \squareforqed              % put the square here
    \parfillskip=0pt           % the line really ends here
    \finalhyphendemerits=0     % ignore a hyphen on the line above
    \endgraf}                  % end the paragraph
  \fi}

\def\blksquare{\rule{2mm}{2mm}}
\def\qedsymbol{\blksquare}
\newcommand{\bg}[1]{\medskip\noindent{\bf #1}}
\newcommand{\ed}{{\hfill\qedsymbol}\medskip}

\newcommand{\R}{\ensuremath{\mathbb R}}

\newcommand{\comment}[1]{}
 {}% the variant that creates the long version

\newcommand{\junk}[1]{}

\newlength{\tmp} \newlength{\lpsx} \newlength{\lpsy} \newlength{\upsx} \newlength{\upsy}

\newcommand{\numbidders}{n}
\newcommand{\numitems}{m}

\begin{document}

\title{Limits of Efficiency in Sequential Auctions}

\author{
Michal Feldman\inst{1}
\and
Brendan Lucier\inst{2}
\and
Vasilis Syrgkanis\inst{3}
}
\institute{
Hebrew Univsersity\\
\email{mfeldman@huji.ac.il}
\and
Microsoft Research New England\\
\email{brlucier@microsoft.com}
\and
Dept of Computer Science, Cornell University\\
\email{vasilis@cs.cornell.edu} 
}

\maketitle

\begin{abstract}
% !TEX root=limits-of-efficiency.tex

We study the efficiency of sequential first-price item auctions at (subgame perfect) equilibrium. This auction format has recently attracted much attention, with previous work establishing positive results for unit-demand valuations and negative results for submodular valuations. This leaves a large gap in our understanding between these valuation classes. In this work we resolve this gap on the negative side. In particular, we show that even in the very restricted case in which each bidder has either an additive valuation or a unit-demand valuation, there exist instances in which the inefficiency at equilibrium grows linearly with the minimum of the number of items and the number of bidders. Moreover, these inefficient equilibria persist even under iterated elimination of weakly dominated strategies. Our main result implies linear inefficiency for many natural settings, including auctions with gross substitute valuations, capacitated valuations, budget-additive valuations, and additive valuations with hard budget constraints on the payments. Another implication is that the inefficiency in sequential auctions is driven by the maximum number of items contained in any player's optimal set, and this is tight. For capacitated valuations, our results imply a lower bound that equals the maximum capacity of any bidder, which is tight following the upper-bound technique established by Paes Leme et al. \cite{PaesLeme2012}.
\end{abstract}

\section{Introduction}
% !TEX root=limits-of-efficiency.tex
Consider the following natural auction setting.  An auction house has a number of items that are offered for sale in an auction on a particular day.  To orchestrate this, the auction house publishes a list of the items to be sold and the order in which they will be auctioned off.  The items are then sold one at a time in the given order.  A group of bidders attends this session of auctions, with each bidder being allowed to participate in any or all of the single-item auctions that will be run throughout the day.  Since the auctions are run one at a time, in sequence, this format is referred to as a sequential auction.

This way of auctioning multiple items is prevalent in practice, due to its relative simplicity and transparency.  It also arises naturally in electronic markets, such as eBay, due to the asynchronous nature of the multiple single-item auctions that are executed on the platform.  A natural question, then, is how well such a sequential auction performs in practice.  Note that while the auction of a single item is relatively simple, equilibria of the larger game may be significantly more complex.  For instance, a bidder who views two of the items as substitutes might prefer to win whichever sells at the lower price, and hence when bidding on the first item he must look ahead to the anticipated outcome of the second auction.  What's more, the sequential nature of the mechanism implies that the outcome of one auction can influence the behavior of bidders in subsequent auctions.  This gives rise to complex reasoning about the value of individual outcomes, with the potential to undermine the efficiency of the overall auction.

%Studying the theoretical properties of simple auctions that arise in practice can give useful guidelines of how and if auction markets should be redesigned. Sequential single item auctions are admittedly one of the simplest procedures for selling a set of items, especially when these items are owned by different sellers. Sequential auctions can arise naturally, in electronic markets like eBay, in art auctions like Sotheby's and in government spectrum auctions.

In this work we study the efficiency of sequential single-item first-price auctions, where items are sold sequentially using some predefined order and each item is sold by means of a first-price auction.  We study the efficiency of outcomes at subgame perfect equilibrium, which is the natural solution concept for a dynamic, sequential game.  Theoretical properties of these sequential auctions have been long studied in the economics literature starting from the seminal work of Weber \cite{Weber2000}.  However, most of the prior literature has focused on very restricted settings, such as unit-demand valuations, identical items, and symmetrically distributed player valuations.  The few exceptions that have attempted to study equilibria when bidders have more complex valuations tend to have other restrictions, such as a very limited number of players or items \cite{Gale2001,Rodriguez2009,Bae2009,Bae2008}. 
Much of the difficulty in studying these auctions under complex environments and/or valuations stems from the inherent complexity of the equilibrium structure, which (as alluded to above) can involve complex reasoning about future auction outcomes.  %inability to explicitly characterize the equilibrium structure in such settings.

Paes Leme et al. \cite{PaesLeme2012} and Syrgkanis and Tardos \cite{Syrgkanis2012a} circumvented this difficulty by performing an indirect analysis on efficiency using the price-of-anarchy framework.  They showed that when bidders have unit-demand valuations (UD), items are heterogeneous, and bidders' valuations are arbitrarily asymmetrically distributed, then the social welfare at every equilibrium is a constant fraction of the optimal welfare. Syrgkanis and Tardos \cite{Syrgkanis2013} extended this result to no-regret learning outcomes and to settings with budget constraints. On the negative side, Paes Leme et al. \cite{PaesLeme2012} showed that this result does not extend to submodular valuations (SM): there exists an instance with submodular valuations where the unique ``natural" subgame perfect equilibrium leads to inefficiency that increases linearly with the number of items, even for a constant number of bidders.

The above results leave a large gap between the positive regime (unit-demand bidders) and the negative (submodular bidders).
%
%\mfcomment{Rewrite what follows: arrange it such that it is clear that there is an hierarchy here.}
Many natural and heavily-studied classes of valuations fall in the range between UD and SM valuations.  Among them are the following, arranged roughly from most to least general:
\begin{itemize}
\item \emph{Gross-substitutes valuations (GS):}
A valuation satisfies the gross-substitutes valuation property if, whenever the cost of one item increases, this cannot reduce the demand for another item whose price did not increase.
\item \emph{$k$-capacitated valuations ($k$-CAP):} Each player $i$ has a capacity $k_i \leq k$ and a value for each item; the value for a set of items is then the value of the $k_i$ highest-valued items in the set.
\item \emph{Budget-additive valuations (BA):} The value of a player $i$ is additive up to a player-specific budget $B_i$ and then remains constant.
\end{itemize}
The class of GS valuations is motivated by the fact that it is (in a certain sense) the largest class of valuations for which a Walrasian equilibrium is guaranteed to exist \cite{Gul1999}, and a Walrasian equilibrium, if exists, is always efficient (see, e.g., \cite{Nisanagt}).
It is known that every $k$-capacitated valuation satisfies gross substitutes \cite{Cohen11}.  Moreover, every gross substitutes valuation is submodular \cite{Lehmann2001}, and it is easy to see that unit-demand valuations are precisely $1$-capacitated valuations.  We therefore have UD $\subset$ $k$-CAP $\subset$ GS $\subset$ SM.  The set of budget-additive valuations is incomparable to UD, $k$-CAP, and GS, but it is known that BA $\subset$ SM.

We ask: \emph{for which of the above classes does the sequential first-price auction obtain a constant fraction of the optimal social welfare at equilibrium?} In this work we show that the answer to the above question is \emph{none of them}.

Specifically, we show that for the case of gross substitutes valuations and for budget additive valuations, the inefficiency of equilibrium can grow linearly with the number of items and the number of players.  Thus, even for settings in which a Walrasian equilibrium is guaranteed to exist, an auction that handles items sequentially cannot find an approximately optimal outcome at equilibrium.  For the case of $k$-capacitated valuations, we show that the inefficiency can be as high as $k$.  This bound of $k$ is tight, following the upper bound established by \cite{PaesLeme2012}.

To prove these lower bounds we consider a different, conceptually more restrictive, class of valuations: the union of unit-demand and additive valuations.  We construct an instance in which every bidder has either a unit-demand valuation or an additive valuation, then show that the unique ``natural'' equilibrium for this instance has extremely poor social efficiency.  We then adapt this construction to provide a lower bound for the valuation classes described above.

%Specifically, we present a case where the inefficiency of equilibrium can grow linearly with the number of items and the number of players, where each agent's valuation is either additive (where the value of a set is the sum of the individual values of its items) or unit-demand (where the value of a set is the maximum value of any item in the set).  This negative example immediately implies a negative answer for $k$-demand valuations and gross-subtitutes valuations.  Thus, even for settings in which a Walrasian equilibrium is guaranteed to exist, an auction that handles items sequentially cannot find an approximately optimal outcome at equilibrium.
%Our example can also be modified to apply to the case of budget additive valuations.
%In particular, we identify an instance in which players can have either unit-demand valuations (where the value of a set is the maximum value of any item in the set) or additive valuations (where the value of a set is the sum of the individual values of its items), and where the inefficiency can grow linearly with the number of items and the number of players.
%Our lower bound construction can be also used to show that the linear lower bound carries over to the class of budget-additive valuations.
%In addition, our results imply that the inefficiency that arises in instances with $k$-demand valuations can be as high as $k$.
%This bound of $k$ is tight, following the upper bound established by \cite{PaesLeme2010}.

We also extend our lower bound to apply to one other setting: additive valuations when players have hard budget-constraints on their payments.  This setting falls outside the quasi-linear regime, but is very relevant in the sequential auction setting: for instance, each bidder may arrive at an auction session with only a certain fixed amount of money to spend.  Note that this is different from the BA valuation class, since it does not restrict the value of a player for a set of items, but rather limits the total payment that a player can make.
For this setting, it is known that maximizing welfare is not an achievable goal in most auction settings, as a participant with low budget is necessarily ineffective at maximizing the value of the item(s) she obtains.  Instead, the natural notion of social efficiency is the ``effective welfare,'' in which the contribution of each participant to the welfare is capped by her budget \cite{Syrgkanis2013}.  We show that, even comparing against the benchmark of effective welfare, our negative result also applies to this setting: for additive valuations with hard budget constraints, the inefficiency can grow linearly with the number of items or players.
This is in stark contrast to the setting of \emph{simultaneous} first-price auctions, where it is known that a constant fraction of the optimal effective social welfare occurs at equilibrium for bidders with hard budget constraints, even when valuations are fractionally subadditive \cite{Syrgkanis2013} (where this class falls between submodular and subadditive valuations).

Sequential auctions with additive bidders and hard budget constraints have been studied in only very limited settings in the economics literature and have recently begun to attract the attention of the computer science community \cite{huang12}. Our result shows that if one allows for arbitrary additive valuations, then such an auction process can lead to very high inefficiency.
%Our lower  bound for this instance, is also a reformulation of our main lower bound instance.

All of the negative results described above rely heavily on the fact that items can be sold in an arbitrary order.
This leads naturally to the following \emph{design} question: does there always exists an order on the items that results in better outcomes at a subgame perfect equilibrium?  This can be interpreted as a mechanism design problem, in which the auctioneer wishes to choose the order in which items are sold in order to mitigate the social impact of strategic bidding. We conjecture that a concrete class of item orders (that we propose) always contains a good order that leads to the VCG outcome at equilibrium, for the class of {\em single-valued unit-demand} valuations.  We leave the resolution of this conjecture as an 
%very interesting 
open problem. 
%
%As partial progress toward resolving this question, we propose an order on the items for the special case of {\em single-valued unit-demand} valuations; that is, where each agent is associated with a non-negative value $v_i$, and his value for any item is either $v_i$ or $0$.  We conjecture that, for this class of valuations, selling the items in the specific order that we construct always results in an equilibrium that realizes a fully efficient outcome, and recovers the VCG payments and view this direction.  
%
%We prove this conjecture under a certain technical condition on the valuation profile, but leave the general question open.  We view this direction as partial evidence that the order in which items are sold is crucial for generating good outcomes at equilibrium. It is tantalizing to suspect that there always exists an order of sale that yields a constant fraction of the optimal social welfare, for valuation classes beyond unit-demand; we leave this as a direction for future research.

\subsection{Related Work}
% !TEX root=limits-of-efficiency.tex

Sequential auctions have been long studied in the economics literature. Weber \cite{Weber2000} and
Milgrom and Weber \cite{Milgrom1982a} analyzed first- and second-price sequential
auctions with identical items and unit-demand bidders in an incomplete-information setting and showed that the unique symmetric equilibrium is efficient and the
prices have an upward drift. The behavior of prices in sequential studies was subsequently studied in
\cite{Ashenfelter1989,McAfee1993}. Boutilier el al. \cite{Boutilier99} studies first-price auctions in a setting with uncertainty, and devised a dynamic-programming algorithm for finding the optimal strategies (assuming stationary distribution of others' bids).

The setting of multi-unit demand has also been studied under the complete-information model. Several papers
studied the two-bidder case, where there is a unique
subgame perfect equilibrium that survives the iterated elimination of weakly
dominated strategies (IEWDS) \cite{Gale2001,Rodriguez2009}.  %, but this is not the case for more than two bidders.
Bae et
al. \cite{Bae2009,Bae2008} studied the case of sequential second-price auctions of identical items with two
bidders with concave valuations and showed that the unique
outcome that survives IEWDS achieves a social welfare at least $1-e^{-1}$ of the optimum.
Here we consider more than two bidders and heterogeneous items.

Recently, Paes Leme et al. \cite{PaesLeme2012} analyzed sequential first- and second-price auctions for heterogeneous items
and multi-unit demand valuations in the complete-information setting. For sequential first-price auctions they showed
that when bidders are unit-demand, every subgame perfect equilibrium achieves at least 1/2 of the optimal welfare,
while for submodular bidders the inefficiency can grow with the number of items, even with a constant number of bidders.
The positive results were later extended to the incomplete-information setting in \cite{Syrgkanis2012a} and to no-regret
outcomes and budget-constrained bidders in \cite{Syrgkanis2013}. In this work we close the gap between positive and negative
results and show that inefficiency can grow linearly with the minimum of the number of items and bidders even when bidders
are either additive or unit-demand.

This work can be seen as part of the recent interest line of research on simple auctions. The closest literature to our work is the that of simultaneous item-bidding auctions \cite{Bikhchandani1999,Christodoulou2008,Bhawalkar2011,Hassidim2011,Feldman2013,Syrgkanis2013}, which is the simultaneous counterpart of sequential auction.
In contrast to sequential auctions, in simultaneous item auctions constant efficiency guarantees have been established for general complement-free valuations, even under incomplete-information settings or outcomes that emerge from learning behavior.
We refer to \cite{DiningBidder} for a recent survey on the efficiency of simultaneous and sequential item-auctions.

\section{Model and Preliminaries}
% !TEX root=limits-of-efficiency.tex

We consider settings with $\numbidders$ bidders and $\numitems$ items, where every bidder $i \in [\numbidders]$ has a valuation function $v_i:2^{[\numitems]} \rightarrow \R_{+}$, associating a non-negative real value with every subset of items.
We denote the set of bidders by $[\numbidders]$ and the set of items by $[\numitems]$.
The valuation function is assumed to be monotone (i.e., $v_i(T) \leq v_i(S)$ for every $T \subseteq S$).
An {\em allocation} is a vector $x = (x_1, \ldots, x_n)$, where $x_i$ denotes the set of items allocated to bidder $i$, and such that $x_i \cap x_j = \emptyset$ for every $i \neq j$. 
%and $\bigcup_{i \in [\numbidders]}x_i = [\numitems]$.

\paragraph{Sequential item auctions.}
The auction proceeds in steps, where a single item is sold in every step using a first-price auction.
In every step $t = 1, \ldots, \numitems$, every bidder $i$ offers a bid $b_i(t)$, and the item is allocated to the agent with the highest bid for a payment that equals his bid.
Each bid in each step can be a function of the history of the game, which is assumed to be visible to all bidders.
More formally, a strategy of bidder $i$ is a function that, for every step $t$, associates a bid as a function of the sequence of the bidding profiles in all periods $1, \ldots, t-1$.
The {\em utility} of an agent is defined, as standard, to be his value for the items he won minus the total payment he made throughout the auction (i.e., quasi-linear utility). We will also assume that the bid space is discretized in
small negligible $\delta$-increments, and for ease of presentation we will use $b^+$ to denote the bid $b+\delta$.

This setting is captured by the framework of extensive-form games (see, e.g., \cite{PaesLeme2012}), where the natural solution concept is that of a {\em subgame-perfect equilibrium} (SPE).
In an SPE, the bidding strategy profiles of the players constitute a Nash equilibrium in every subgame.
That is, at every step $t$ and for every possible partial bidding profile $b(1), b(2), \dotsc, b(t-1)$ up to (but not including) step $t$, the strategy profile in the subgame that begins in step $t$ constitutes a Nash equilibrium in the induced (i.e., remaining) game.

\paragraph{Elimination of Weakly Dominated Strategies.}
We wish to further restrict our attention to ``natural'' equilibria, that exclude (for example) dominated overbidding strategies.  We therefore consider a natural and well-studied refinement of the set of subgame perfect equilibria: those that survive iterated elimination of weakly dominated strategies (IEWDS).  A strategy $s$ is \emph{weakly dominated} by a strategy $s'$ if, for every profile of other players' strategies $s_{-i}$, we have $u_i(s,s_{-i}) \leq u_i(s', s_{-i})$, and moreover there exists some $s_{-i}$ such that $u_i(s, s_{-i}) < u_i(s', s_{-i})$.  Roughly speaking, under IEWDS, each player removes from her strategy space the set of all weakly dominated strategies.  This removal may cause new strategies to become weakly dominated for a player, which are then removed from her strategy space, and so on until no weakly dominated strategies remain.  We defer a formal definition of IEWDS to Appendix \ref{app:iewds}.  %The following definition makes this concept more precise.

%\begin{definition}
%Given an $n$-player game defined by strategy sets $S_1, \dotsc, S_n$ and utilities $u_i \colon S_1 \times \dotsc \times S_n \to \mathbb{R}$ we define a \emph{valid procedure for eliminating weakly dominated strategies} as a sequence $\{S_i^t\}$ such that for each $t$ there is an $i$ such that $S_j^t = S_j^{t-1}$ for $j \neq i$, $S_i^t \subseteq S_t^{t-1}$, and for all $s_i \in S_i^{t-1} \backslash S_i^t$ there is some $s_i' \in S_i^t$ such that $u_i(s_i', s_{-i}) \geq u_i(s_i, s_{-i})$ for all $s_{-i} \in \prod_{j \neq i}S_j^t$ and the inequality is strict for at least one $s_{-i}$.  We say that a strategy profile $s$ survives iterated elimination of weakly dominated strategies (IEWDS) if, for any valid procedure $\{S_i^t\}$, $s_i \in \cap_t S_i^t$.
%\end{definition}

We will focus on subgame perfect equilibria of sequential first-price item auctions that survive IEWDS.  It is shown in \cite{PaesLeme2012} that there always exists such an equilibrium. 
%A characterization of such equilibria is also given in \cite{PaesLeme2012}.  
We note one necessary property of an equilibrium satisfying IEWDS: in every subgame beginning at a time $t = m$ (i.e., when the last item is being sold), for every possible bidding history up to that round, each player will bid no more than his marginal value for the final item.  In other words, no player can credibly threaten to overbid on the last item for sale.

\paragraph{Price of anarchy.}
The price of anarchy (PoA) measures the inefficiency that can arise in strategic settings.
The PoA for subgame perfect equilibria is defined as the worst (i.e., largest) possible ratio between the welfare obtained in the optimal allocation and the welfare obtained in any subgame perfect equilibrium of the game.  We note that all of our lower bounds on the price of anarchy will involve ``natural'' equilibria that survive IEWDS.

\section{A Simple Example}\label{sec:example}
% !TEX root=limits-of-efficiency.tex

To develop some intuition regarding the strategic considerations that might take place in sequential auctions, we give a simple example in which one bidder has value for many items (i.e., wholesale buyer) and another bidder has value for only one item (i.e., retail buyer).

In particular, consider a sequence of two auctions for two identical items and two buyers, $A$ and $B$.  Buyer $A$ is a ``wholesale'' buyer, having an additive valuation with a value of $9$ for each of the two items.  Buyer $B$ is a ``retail'' buyer, who wants only one item (unit-demand) and has a value of $5$ for either of the two.
The items are sold sequentially using a first-price auction for each item.

Consider the situation from the perspective of the additive buyer $A$. Thinking strategically and farsightedly, he reasons that if he wins the first auction, then in the second auction he will have to compete with buyer $B$ and will therefore have to pay $5$ dollars to win the second item.
If, however, he lets buyer $B$ win the first item, then buyer $B$ will have no value for the second item and hence the only undominated strategy for buyer $B$ will be to bid $0$ in the second auction, and hence buyer $A$ will win the second item for free.
What must buyer $A$ pay in order to win the first item?
Buyer $B$ knows that if the first item goes to buyer $A$, then buyer $B$ will certainly lose the second item as well; therefore buyer $B$ is willing to pay up to $5$ for the first item.
Therefore, in order to win the first item, buyer $A$ will have to bid at least $5$ in the first auction.

Thus bidder $A$ needs to choose between the following two options:
he can either win both auctions and pay a price of $5$ for each one of them, or let bidder $B$ win the first auction and win only the second auction but pay nothing.
Observe that the first option gives bidder $A$ a utility of $8$ ($=2 \cdot (9-5)$) while the second option gives him a utility of $9$ ($=1 \cdot (9-0)$).
Consequently, bidder
$A$ will choose to forego the first item in order to improve his situation in the second one.
Interestingly, this outcome is socially suboptimal, since the efficient outcome is for bidder $A$ to win both items ---
although bidder $A$ has much more value for the first item than bidder $B$, the first item is allocated to $B$ in equilibrium.

One can also take this example to the extreme where, e.g., bidder $A$'s value is set to
$10-\epsilon$ for each item. In this case the unique subgame perfect equilibrium that survives elimination of dominated strategies is a $4/3$ approximation to the optimal welfare, even though the items are identical (and therefore the inefficiency is irrespective of item ordering).
In the next section we demonstrate that with heterogeneous items, the social welfare of sequential item auctions
at subgame perfect equilibrium can be as low as an $O(m)$ fraction of the optimal social welfare. 

\section{Lower Bound for Additive and Unit-demand Valuations}
% !TEX root = limits-of-efficiency.tex

We now present our main result by providing an instance of a sequential first price auction with unit-demand and
additive bidders, where the social welfare at a subgame-perfect equilibrium that survives IEWDS\footnote{
The equilibrium that we describe
is, in some sense, the unique natural equilibrium: if we were to ask players to submit bids sequentially within each 
auction, rather than simultaneously, then there would be a unique equilibrium (solvable by backward induction), 
which is the equilibrium that we describe.
} achieves social
welfare that is only an $O(\min\{n,m\})$-fraction of the optimal welfare. 
Therefore, our example shows that inefficiency can arise at equilibrium in a robust manner.

\begin{theorem}
\label{thm.poa.bad}
The price of anarchy of the sequential first-price item auctions with additive and unit-demand bidders is $\Omega( \min\{n,m\} )$.  Moreover, this result persists even if we consider only equilibria that survive IEWDS.
\end{theorem}

\textbf{Informal Description.} Before we delve into the details of the proof of Theorem \ref{thm.poa.bad}, we give a high-level idea of the type of strategic manipulations that lead to inefficiency and compare them with the simultaneous auction counterpart of our sequential auction.

Consider an auction instance where two additive bidders have identical values for most of the items for sale, but their valuations differ only on the last few items that are sold.  Specifically, assume that there are two items $Z_1$ and $Z_2$, auctioned last, such that only player $1$ has value for $Z_1$ and only player $2$ has value for $Z_2$. We will refer to these items as the \textit{non-competitive items} and to all other items as the \textit{competitive items}. The additive bidders know that it is hopeless to try to achieve any positive utility from the competitive items on which they have identical interests. The only utility they can ever derive is from the last, non-competitive items on which they don't compete with each other.  If these were the only two players in the auction, then we would obtain the optimal outcome: the two bidders would simply compete on each of the competitive items, with one of them acquiring each competitive item at zero utility.\footnote{In fact, optimality is always achieved when all bidders are additive, in general.}

We now imagine adding unit-demand bidders to the auction in order to perturb the optimality.
Specifically, suppose there is
a unit-demand bidder that has value for the two \textit{non-competitive items}, with the value for item $Z_i$ being slightly less than player $i$'s value for $Z_i$, $i \in \{1,2\}$. This endangers the additive bidders' hopes of getting non-negligible utility, since competition from the unit-demand player may drive up the prices of $Z_1$ and $Z_2$.  The only hope that the additive bidders have is that the unit-demand bidder will have his demand satisfied prior to these final two auctions, in which case the unit-demand bidder would not bother to bid on them. Hence, the two additive bidders would do anything in their power to guide the auction to such an outcome, even if that means sacrificing all the competitive items! This is exactly the effect that we achieve in our construction. Specifically, we create an instance where this competing unit-demand bidder
has his demand satisfied prior to the auctions for $Z_1$ and $Z_2$ if and only if a very specific outcome occurs: the additive bidders don't bid at all on all the competitive items, but rather other small-valued bidders acquire the competitive items instead. These small-valued bidders contribute almost nothing to the welfare, and therefore all of the welfare from the competitive items is lost.

It is useful to compare this example with what would happen if the auctions were run simultaneously, rather than sequentially.  This uncovers
the crucial property of sequential auctions that leads to inefficiency: the \textit{ability
to respond to deviations}. If all auctions happened simultaneously, then the behavior of the additive bidders that we described above could not possibly be an equilibrium: one additive bidder, knowing that his additive competitor bids $0$ on all the
competitive items, would simply deviate to outbid him on the competitive items and get a huge utility. However, because the items are sold sequentially, this deviation cannot be undertaken without consequence: the moment one of the additive bidders deviates to bidding on the competitive items, in all subsequent auctions
the competitor will respond by bidding on subsequent competitive items, leading to zero utility for the remainder of the auctions.  Moreover, this response need not be punitive, but is rather the only rational response once the auction has left the equilibrium path (since the additive bidders know that there is no way to obtain positive utility in subsequent auctions).  Thus, in a sequential auction, an additive player can only
extract utility from at most one competitive item, which is not sufficient to counterbalance the resulting utility-loss due to the increased competition on the last non-competitive item.

\textbf{The Lower Bound.} 
We now proceed with a formal proof of Theorem \ref{thm.poa.bad}.  
Consider an instance with $2$ additive players, $k$ unit-demand players and $k+3$ items.
Denote with $\{a,b\}$ the two additive players and with $\{p_1,\ldots,p_k\}$ the $k$
unit-demand players. Also denote the items with $\{I_1, \hdots, I_k, Y, Z_1, Z_2\}$. The
valuations of the additive players are represented by the following table of $v_{ij}$, where
$\epsilon > 0$ is an arbitrarily small constant:

\begin{center}
\begin{tabular}{ c || c | c | c | c | c | c  }
    & $I_k$ & $\hdots$ & $I_1$ & $Y$ & $Z_1$ & $Z_2$ \\
  \hline
  $a$ & $1+\epsilon$ & $\hdots$ & $1+\epsilon$ & $0$ & $ 10$ & $0$ \\
  $b$ & $1$ & $\hdots$ & $1$ & $0$ & $0$ & $ 10$ \\
\end{tabular}
\end{center}

In addition the unit-demand valuations for the $k$ players are given by
the table of $v_{ij}$ that follows (an empty entry corresponds to a $0$ valuation), though now a valuation of a player when getting a set $S$ is
$\max_{j\in S}v_{ij}$:

\begin{center}
\begin{tabular}{ c || c | c | c | c | c | c | c | c | c }
    & $I_k$ & $I_{k-1}$ & $I_{k-2}$ & $\hdots$ & $I_2$ & $I_1$ & $Y$ & $Z_1$ & $Z_2$ \\
  \hline
  $p_0$ &  &  &  &  $\hdots$ &  &  & $10-\epsilon$ & $10-\epsilon$ & $10-\epsilon$ \\
  $p_1$ &  &  &  &  $\hdots$ &  & $\delta_1$ & $10$ &  &  \\
  $p_2$ &  &  &  &  $\hdots$ & $\delta_2$ & $\delta_2$  &  &  &  \\
  \ldots \\
  $p_{k-1}$ &  & $\delta_{k-1}$ & $\delta_{k-1}$ & $\hdots$  &  &  &  & \\
  $p_k$ & $\delta_k$ & $\delta_k$ &  & $\hdots$  &  &  &  & \\
\end{tabular}
\end{center}
The constants $\delta_1, \dotsc, \delta_k$ are chosen to satisfy
the following condition:
\begin{equation}\label{eqn:assum}
\delta_k > \delta_{k-1} > \hdots > \delta_2 > \delta_1 > \epsilon
\end{equation}
Note that, by taking $\epsilon$ to be arbitrarily small, we can take each $\delta_i$ to
be arbitrarily small as well.

In the optimal allocation, player $a$ gets all the items $I_1, \hdots, I_k$ and $Z_1$,
player $b$ gets $Z_2$ and player $p_1$ gets $Y$. The resulting social welfare
is $k(1+\epsilon) + 30$. 
We assume that the auctions take place in the order depicted in the valuation tables:
$\{I_k, \ldots, I_1, Y, Z_1, Z_2\}$. 
We will show that there is a subgame perfect
equilibrium for this auction instance such that the unit-demand players win all the items $I_1,\hdots, I_k$.
Specifically, player $p_i$ wins item $I_i$, player $a$ wins $Z_1$, player $b$ wins
$Z_2$, and player $p_0$ wins $Y$, resulting in a social welfare of
$30 - \epsilon + \sum_{i = 1}^k \delta_i$.  Taking $\delta$ sufficiently small, this welfare
is at most $31$.  This will establish that the price of anarchy for this instance is at least
$\frac{k(1+\epsilon)+30}{31} = O(k)$, establishing Theorem \ref{thm.poa.bad}.
%
%since we assume that the delta's are very small compared to $1$.
Furthermore, we will show that this subgame perfect equilibrium is \emph{natural},
in the sense that it survives iterated deletion of weakly dominated strategies.

The intuition is the following: 
after the first $k$ auctions have been sold, player $p_0$ has to decide if he will target (and win) item $Y$, 
or if he will instead target items $Z_1$ and/or $Z_2$. If he targets item $Y$, he competes with player $p_1$ and afterwards
lets players $a$ and $b$ win items $Z_1, Z_2$ for free. This decision of player
$p_0$ depends on whether player $p_1$ has won item $I_1$, which in turn depends on
the outcomes of the first $k-1$ auctions. In
particular, player $p_1$ can win item $I_1$ only if player $p_2$ has won item $I_2$. In turn,
$p_2$ can win $I_2$ only if $p_3$ has won item $I_3$ and so on. Hence, it will turn out that 
in order for $p_0$ to want to target item $Y$,
it must be that each item $I_i$ is sold to bidder $p_i$.  Thus, if either
player $a$ or $b$ acquires any of the items  $I_1, \hdots,
I_k$, they will be guaranteed to obtain low utility on items $Z_1$ and $Z_2$.  This will
lead them to bidding truthfully on all subsequent $I_i$ auctions, leading to a severe drop
in utility gained from future auctions.
%deriving very little utility from any auction in the future.

In the remainder of this section, we provide a more formal analysis of the equilibrium 
in this auction instance.  We
begin by examining what happens in the last three auctions of $Y,Z_1$ and $Z_2$, conditional
on the outcomes of the first $k$ auctions.
We first examine the outcome of auctions $Y,Z_1,Z_2$ conditional on the outcome of auction
$I_1$:

\begin{itemize}
\item Case 1: $p_1$ has won $I_1$

Player $p_1$ has marginal value of $10-\delta_1$ for item $Y$.  Hence, he is willing
to bid at most $10-\delta_1$ on item $Y$.

Player $p_0$ knows that if he loses $Y$ then in the subgame
perfect equilibrium in that subgame he will bid $10-\epsilon$ on $Z_1$ and $Z_2$ and lose.
Thus he expects no utility from the future if he loses $Y$. Thus he is willing to
pay at most $10-\epsilon$ for item $Y$.

Since by assumption \eqref{eqn:assum} $\delta_1>\epsilon$, player $p_0$ will win $Y$ at
a price of $10 - \delta_1$.  
%$\delta_1-2\epsilon$. 
Then players $a,b$ will win $Z_1$ and $Z_2$ for free. Thus
the utilities in this case from this subgame are: $u(a)=10$, $u(b)=10$, $u(p_0)=\delta_1-\epsilon$, $u(p_1)=0$.

\item Case 2: $p_1$ has lost $I_1$

Player $p_1$ has marginal value of $10$ for item $Y$. Hence, he is willing to bid at most $10$ on item $Y$.

Player $p_0$ performs the exact same thinking as in the previous case and thereby is willing to bid at most
$10-\epsilon$ for item $Y$.

Thus in this case $p_1$ will win item $Y$ at a price of $10-\epsilon$. Then, as predicted, $p_0$ will 
bid $10-\epsilon$ on $Z_1$ and $Z_2$ and lose. Thus the utilities
of the players in this case are: $u(a)=\epsilon$, $u(b)=\epsilon$, $u(p_0)=0$, $u(p_1)=\epsilon$.

\end{itemize}

Now we focus on the auction of item $I_1$. As was explained in Paes Leme et al. \cite{PaesLeme2012}
this auction will be an auction with externalities where each player has a different utility for
each different winner outcome. This utilities can be concisely expressed in a table of $v_{ij}$'s
where $v_{ij}$ is the value of player $i$ when player $j$ wins. The only players that potentially have any
incentive to bid on item $I_1$ are $a,b,p_0,p_1,p_2$. The following table summarizes their
values for each possible winner outcome of auction $I_1$ as was calculated in the previous case-analysis
(we point that in the diagonal we also add the actual value that a player acquires from item $I_1$ to his future utility
conditional on winning $I_1$) .

$$
[v_{ij}]=\begin{tabular}{c|ccccc}
 ~  & $a$ & $b$ & $p_0$ & $p_1$ & $p_2$\\
\hline
$a$ & $1+2\epsilon$ & $\epsilon$ & $\epsilon$ & $10$ & $\epsilon$\\
$b$ & $\epsilon$ & $1+\epsilon$ & $\epsilon$ & $10$ & $\epsilon$ \\
$p_0$ & $0$ & $0$ & $0$ & $\delta_1-\epsilon$ & $0$ \\
$p_1$ & $\epsilon$ & $\epsilon$ & $\epsilon$ & $\delta_1$ & $\epsilon$ \\
$p_2$ & $0$ & $0$ & $0$ & $0$ & $\delta_2\cdot {\bf 1}_{\text{hasn't won } I_2}$\end{tabular}
$$

For example, player $a$ obtains utility $10$ if player $p_1$ wins item $I_1$.
We see from the table that, at this auction, everyone except $p_2$ achieves their maximum 
value when $p_1$ wins the auction. Player
$p_2$ has value for winning the auction only if he hasn't won $I_2$. In addition, since
$\delta_2>\delta_1$, if $p_2$ hasn't won $I_2$ then he can definitely outbid $p_1$ on $I_1$ 
and therefore $p_1$ has no chance of winning the auction of $I_1$. As we now show, this 
implies that there is a unique equilibrium of the auction conditioning on whether or not $p_2$ has won $I_2$:
\begin{itemize}
\item Case 1: If $p_2$ has won $I_2$ then he has no value for $I_1$. There exists an equilibrium in
undominated strategies where and all players $a,b,p_0,p_2$ will bid $0$, while $p_1$ bids $0^+$.
In fact this is in some sense the most natural equilibrium since it yields the highest utility for
$a$ and $b$. In this case the utility of the players from auctions $I_1$ and onward will be:
$u(a)=10$, $u(b)=10$, $u(p_0)=\delta_1-\epsilon$, $u(p_1)=\delta_1$, $u(p_2)=0$.

\item Case 2: If $p_2$ has lost $I_2$, then he has value of $\delta_2>\delta_1$ for $I_1$. Hence,
$p_1$ has no chance of winning item $I_1$. Thus, the unique equilibrium that survives elimination of weakly
dominated strategies in this case is for player $a$ to bid $1^+$, for player $b$ to bid $1$,
for player $p_0$ to bid $0$, for player $p_1$ to bid $\delta_1-\epsilon$ and for player $p_2$ to bid $\delta_2$.  In this case the utility of the players from auctions $I_1$ and on will be: $u(a)=2\epsilon$, $u(b)=\epsilon$, $u(p_0)=0$, $u(p_1)=\epsilon$, $u(p_2)=0$.
\end{itemize}

Using similar reasoning we deduce that player $p_i$ can win $I_i$ only if $p_{i-1}$ has won $I_{i-1}$.
If at any point some $p_i$ does not win $I_i$ then players $a$ and $b$ know that from that point onward no $p_j$
can win auction $I_j$, and therefore they will get only utility $\epsilon$ from $Z_1, Z_2$. Thus there will 
be no reason for players $a$ and $b$ to
allow unit-demand players to continue to win items, and thus the only equilibrium strategies 
from that point on will be for $a$ to
bid $1^+$ on each of $I_i$ and $b$ to bid $1$. This will lead to player $a$ to get utility
$O(\epsilon)$ from each auction for items $I_{i-1}, \dotsc, I_2$, and player $b$ to get no utility from 
these auctions. Thus, at any
point in the auction, it is an equilibrium for players $a$ and $b$ to allow the unit
demand player $p_i$ to win auction $I_i$ conditional on the fact that they have allowed all
previous unit-demand bidders to win. In particular, in the first auction, it is an equilibrium for players
$a$ and $b$ to allow player $p_k$ to win.  We conclude that the strategy profile we described is
a subgame perfect equilibrium for this auction instance.  This completes the proof of 
Theorem \ref{thm.poa.bad}

Finally, as discussed throughout our analysis, the equilibrium described above survives IEWDS.  
The reason is that, for every item $k$ and
bidder $i$, the proposed equilibrium strategy for bidder $i$ does not require that he bid more than
his value for item $k$ less his utility in the continuation game subject to not winning item $k$.  As
discussed in Paes Leme et al. \cite{PaesLeme2012}, this property guarantees that no player
is playing a weakly dominated strategy.

%\blcomment{How much more should we say about IEWDS?  Is it self-evident, or is it worth giving
%a full proof?  Something to think about if we decide to write up the construction in this
%section more formally.}
%\vscomment{I think that more formal analysis might be tiring. I thought the above was formal enough. I don't think we should say much more about IEWDS.}

\section{Extensions of the Lower Bound}
% !TEX root=limits-of-efficiency.tex

We now provide some reinterpretations and extensions of our lower bound from the previous section, to show that linear inefficiency can occur under several important classes of valuations.
%that show that show that inefficiency can grow linearly with the number of items for
%several classes of valuations.

\paragraph{Gross Substitutes.} Since the class of gross substitutes valuations includes all additive and unit-demand valuations, the example from the
previous section immediately implies a linear price of anarchy for gross substitutes valuations.

\paragraph{Budget-Additive.} A valuation is budget additive if it can be written in the form $v(S) = \max\left\{B,\sum_{j\in S}v_j\right\}$. As it turns out, in
the example in the previous section all valuations are budget additive.  The additive players can be thought of as having infinite
budget.  Each of the unit-demand players $p_i$ for $i\in [2,k]$ can be thought as budget-additive with a budget of $\delta_i$ and value
$\delta_i$ for items $I_{i}$ and $I_{i+1}$ and $0$ for everything else.  Player $p_1$ has budget of $10$ and additive value of $\delta_1$ for $I_1$, $10$ for $Y$
and $0$ for everything else. Player $p_0$ has budget $10-\epsilon$ and additive value of $10-\epsilon$ for each of $Y, Z_1, Z_2$ and $0$ for everything else.
Therefore the analysis in the previous section holds even for budget-additive valuations.
%\bldelete{The key reason, that we could adapt the example to budget-additive valuations
%is that all unit-demand players have their highest valued item (among the items that they have positive value) auctioned last.
%}

%\bldelete{
%More concretely, if we focus on the crucial auction of $Y$, then we have that if player $p_1$ has already won $I_1$ then he is willing to bid at most
%$10-\delta_1$ on $Y$, since his budget becomes binding. Additionally, player $Y$ is willing to pay at most $10-\epsilon$. Since $\delta_1>\epsilon$ we have that player $Y$ will win the auction. After that he has no extra marginal value for any subsequent items since he has exhausted his valuation budget and the
%additive bidders will win for a price of $0$ both $Z_1$ and $Z_2$. On the other hand if player $p_1$ has lost $I_1$ then he is willing
%to bid $10>10-\epsilon$ on $Y$ and thereby, $p_0$ will lose $Y$. Subsequently, $p_0$ will bid $10-\epsilon$ on $Z_1$ and $Z_2$ and lose each one, leading to an almost $0$ utility for the additive players. The remainder of the analysis is the same as in the previous section.
%}

\paragraph{Additive valuations with budget constraints on payments.} We show that the same analysis can be applied to a setting in which each player $i$ has
an additive valuation %: i.e. $v(S)=\sum_{j\in S} v_j$. Additionally, he
as well as a hard budget constraint $B_i$ on his payment.  That is, his utility
is quasi-linear as long as his payment is below $B_i$, but becomes minus infinity if he pays more than $B_i$. Formally, if a player $i$
receives a set $S$ and pays total price $p$ then his utility $u_i(S,p)$ is $ v_i(S) - p$ if $p \leq B_i$, or $-\infty$ otherwise.
%\begin{equation}
%u_i(S,p)=\begin{cases}
%v_i(S) - p & \text{~if~}p\leq B_i\\
%-\infty & \text{~o.w.~}
%\end{cases}
%\end{equation}

We will adapt the example from the previous section to the setting of budget constraints in a manner similar to the case of budget-additive valuations.
%the
%Similarly to the budget-additive case we will try to simulate the behavior of the unit-demand players, with additive players with hard budget constraints.
Specifically, we set the budgets of the players as in the budget-additive case described above, but we treat them as payment budgets rather than a cap on valuations.

We need to be slightly careful in our analysis under this adaptation, since it doesn't only matter whether a player won or lost an item, but also at which price. Specifically,
the equilibrium will alter slightly. The additive bidders, apart from letting bidder $p_i$ win $I_i$, will also have to make him pay enough
so that he has no remaining budget with which to win the subsequent item $I_{i+1}$.

For player $p_0$, we know that his budget is indeed almost exhausted at auction $Y$ whenever he wins, since player $p_1$ has a substantial value. Thus
for auction $Y$ no change in the equilibrium analysis takes place. However, when examining auction $I_1$, if we consider the same equilibrium as in
the previous section, then player $p_i$ pays nothing and thus still has all his budget to bid on $Y$ and win it. It is in the interest of the
additive bidders to ensure that $p_1$ not only wins, but also pays at least $\epsilon$, so that he doesn't have enough budget to win item $Y$. 

Player $p_1$ knows that if he loses the auction for item $I_1$ then he can use his budget to get utility of $\epsilon$ from winning $Y$. If he wins $I_1$ for a price of $t\geq \epsilon$
then he gets no utility from the future and instead gets a utility of $\delta_1-t$ from winning $I_1$. Assuming that $\delta_1> 2\epsilon$, player
$p_1$ is willing to pay more than $\epsilon$ to win auction $I_1$. Thus, if we assume $\delta_1 > 2\epsilon$, the additive players can bid enough on item $I_1$ that
player $p_1$ will win it at some price above $\epsilon$, which will then result in $p_0$ winning $Y$ and the additive bidders getting utility $10$ from $Z_1$ and $Z_2$. A similar analysis holds for the auction of each item $I_i$, for $i\in [2,k]$: the additive players need to make sure that each bidder $p_i$ wins $I_i$, and also pays enough so that he doesn't have enough budget
to tilt player $p_{i+1}$ on getting his next item rather than $I_{i+1}$. However, observe that if player $p_i$ loses auction $I_i$, then
subsequently the additive players will switch to winning all the remaining items, since there is no hope to make the unit-demand bidders win their items; so it is in the interest of each player $I_i$ to accept any price up to $\delta_{i}$ and therefore the additive players can completely exhaust his budget. With this change in the equilibrium
strategies, our analysis in the previous section carries over, and we conclude that the price of anarchy in this instance is $\Omega(k)$.

\section{The Impact of Item Ordering}
\label{sec:ordering}
% !TEX root=limits-of-efficiency.tex
Our lower bound establishes that if items are sold sequentially, then arbitrarily inefficient outcomes can result at equilibrium even when all agents have gross substitutes valuations.  The constructions depend on the items being sold in an arbitrary order.  A natural question arises: does there always exist an order over the items such that the resulting outcome is efficient, or approximately efficient?

In this section we discuss this problem in the context of unit-demand bidders. Recall that, for unit-demand bidders, selling items in an arbitrary order always results in an outcome that achieves at least half of the optimal social welfare. Additionally, it is known by \cite{PaesLeme2012} that if any order is allowed then the unique subgame-perfect equilibrium that survives IEWDS 
can be inefficient, achieving only a $3/2$-approximation. This lower bound of $3/2$ holds even for the special case of single-valued unit-demand bidders, where each player has a single value $v_i$ for getting one item from some interest set $S_i$.  We conjecture that, for the case of single-valued unit-demand bidders, if the auctioneer can choose the order in which the objects are sold, then it is possible to recover the optimal welfare at all natural equilibria.  Indeed, we make a stronger conjecture: there exists an order in which the VCG outcome (allocation and payments) occurs at equilibrium.
%is consider the problem of the efficiency of sequential auctions in settings with single-valued unit-demand bidders. 
%In this setting, there are $m$ items for sale and $n$ agents, where agent $i$ is associated with a value $V_i$ and a subset of the items $S_i$, and for every subset $T \subseteq [m]$, $v_i(T)=V_i$ if $T \cap S_i \neq \emptyset$ and $v_i(T)=0$ otherwise.

%\blcomment{We actually want to say that the VCG outcome is the only natural equilibrium.  Vasilis, is the solution concept that eliminates all equilibria but the one we care about easy to describe?  Is it just iterated deletion of dominated strategies?}
%
%\blcomment{Should we make the conjecture only for single-valued unit demand?  How strongly do we believe it holds for general unit-demand?}

\begin{conjecture}
\label{conj:order}
For every instance of single-valued unit-demand bidders, there exists an order over the items such that the corresponding sequential auction admits a subgame perfect equilibrium that survives IEWDS and that replicates the VCG outcome.
\end{conjecture} 
Observe that such a result cannot hold for both additive and unit-demand bidders as is portrayed by our simple example in Section \ref{sec:example}, where all items are
identical and hence, under any ordering, the unique subgame-perfect equilibrium that survivies IEWDS is inefficient. Our conjecture also stems from the fact that for
the case of single-valued unit-demand bidders the optimization problem is a matroid optimization problem. It is known by \cite{PaesLeme2012} that a form 
of sequential cut auction for matroids always leads to a VCG outcome. The difference is that sequential item-auctions do not correspond to auctions across
cuts of the matroid. However, it is feasible that under some ordering the same behavior as in a sequential cut auction will be implemented.

As progress toward this conjecture, we will present a subset of item orderings, the \emph{augmenting path orderings}, which we believe always contains an ordering that satisfies Conjecture \ref{conj:order}.  For instance, we show in Appendix \ref{app:conj} that the $3/2$ lower bound of \cite{PaesLeme2012} breaks if we only allow augmenting path orderings. 
%We will then show an example that establishes that not all augmenting path orderings satisfy the conjecture and 
We leave open the question of whether one of these orderings always yields a VCG outcome.

\subsection{A Class of Orderings}

Consider a profile of single-valued unit-demand valuations.  Let $x$ denote the VCG allocation (i.e., $x_i$ is the item allocated to bidder $i$).  We also write $x^{(-i)}$ to denote the VCG allocation when bidder $i$ is excluded.  For each $i$, the allocations $x$ and $x^{(-i)}$ define a directed bipartite graph between players and objects, where there is an edge between player $k$ and item $j$ if $x_k^{(-i)} = j$ but $x_k \neq j$, and there is an edge from item $j$ to player $k$ if $x_k^{(-i)} \neq j$ but $x_k = j$.  It is known that, for each player $i$, this graph is always a directed path from player $i$ to some other player $k$; this is the \emph{augmenting path for player $i$} and player $k$ is the \emph{price setter} of player $i$, i.e. the VCG price of player  $i$ is $v_k$. With no loss of generality we assume that every player has a price setter $k$.

Given a welfare-optimal matching $\pi$, that matches each player $i$ to an item $\pi(i)$, consider the following forest construction. Consider all price setters in decreasing value order. For each price setter $k$, we will create a tree and add it to the forest, as follows.  Consider all the items that are in the interest set of $k$, $S_k$, that are
not yet in the forest.  Add each such item to the tree as a child of player $k$.  Next, from each such item $j$, consider its optimally matched player $\pi^{-1}(j)$ and add this player to the tree as a child of $j$.  For each player $i$ that was added, consider all items that are in the interest set of $i$, $S_i$, that are not yet in the forest, and add each of these items to the tree as a child of $i$.  We continue this process, which is essentially a breadth-first traversal of the set of items, until there is no new item to be added. 

The above process creates a forest that contains a node for each item, for each player that is allocated an item in the optimal allocation, and for each price setter. Additionally, each player
belongs to the tree rooted at his price setter and his unique path in the tree to the price setter is an augmenting path in the initial bipartite graph. 
The reasoning is as follows: each tree contains all possible alternating paths ending at the price-setter, except alternating paths that contain items and players who have been included in the tree of a price setter with larger value. Since a player's price setter is the largest unallocated player with which he
is connected, through an alternating path, the claim follows.
%
%It is also known that the union of all augmenting paths, over all players, forms a directed forest (with edges pointing from the leaves to the root of each tree).  \blcomment{Todo: give proofs of these statements, or a citation}.  We refer to this as the augmenting path graph, which we will denote by $G$.

We will refer to the above forest as the \emph{augmenting path graph} $G$. Given an augmenting path graph $G$, a \emph{post-order item traversal of $G$} is a depth-first, post-order traversal of the nodes of $G$, restricted to the nodes corresponding to items and rooted at price setters.  Note that this is an ordering over the items in the auction.  We also assume that trees are traversed in decreasing order of price-setters. Also note that this order is not necessarily unique, as it does not specify the order in which the children of a given node should be traversed.

\begin{defn}
The set of \emph{augmenting path orderings} of the items is the set of orderings corresponding to post-order item traversals of $G$.
\end{defn}

Our (refined) conjecture is that, for every instance of single-valued unit-demand bidders, there exists an augmenting path ordering such that the corresponding sequential auction admits a subgame perfect equilibrium that replicates the VCG outcome.  As an example, we show in Appendix \ref{app:conj} that this conjecture holds for the $3/2$ lower bound example from \cite{PaesLeme2012}.  We also show in Appendix \ref{app:cex} that it is not true that \emph{all} augmenting path orderings lead to efficient outcomes at equilibrium: there are examples in which multiple augmenting path orderings exist, and some orderings lead to inefficient outcomes at equilibrium.

\bibliographystyle{plain}
\bibliography{simple-auctions}

\begin{thebibliography}{10}

\bibitem{Ashenfelter1989}
Orley Ashenfelter.
\newblock {How auctions work for wine and art}.
\newblock {\em The Journal of Economic Perspectives}, 3(3):23--36, 1989.

\bibitem{Bae2008}
Junjik Bae, Eyal Beigman, Randall Berry, Michael Honig, and Rakesh Vohra.
\newblock {Sequential Bandwidth and Power Auctions for Distributed Spectrum
  Sharing}.
\newblock {\em IEEE Journal on Selected Areas in Communications},
  26(7):1193--1203, September 2008.

\bibitem{Bae2009}
Junjik Bae, Eyal Beigman, Randall Berry, Michael~L. Honig, and Rakesh Vohra.
\newblock {On the efficiency of sequential auctions for spectrum sharing}.
\newblock {\em 2009 International Conference on Game Theory for Networks},
  pages 199--205, May 2009.

\bibitem{Bhawalkar2011}
Khsipra Bhawalkar and Tim Roughgarden.
\newblock Welfare guarantees for combinatorial auctions with item bidding.
\newblock In {\em SODA}, 2011.

\bibitem{Bikhchandani1999}
Sushil Bikhchandani.
\newblock Auctions of heterogeneous objects.
\newblock {\em Games and Economic Behavior}, 26(2):193 -- 220, 1999.

\bibitem{Nisanagt}
Liad Blumrosen and Noam Nisan.
\newblock chapter Combinatorial Auctions.
\newblock Camb. Univ. Press, '07.

\bibitem{Boutilier99}
Craig Boutilier, Moises Goldszmidt, and Bikash Sabata.
\newblock {Sequential Auctions for the Allocation of Resources with
  Complementarities}.
\newblock In {\em IJCAI-99: Proceedings of the Sixteenth International Joint
  Conference on Artificial Intelligence}, pages 527--534, 1999.

\bibitem{Christodoulou2008}
George Christodoulou, Annam\'{a}ria Kov\'{a}cs, and Michael Schapira.
\newblock Bayesian combinatorial auctions.
\newblock In {\em ICALP}, 2008.

\bibitem{Cohen11}
E.~Cohen, M.~Feldman, A.~Fiat, H.~Kaplan, and S.~Olonetsky.
\newblock Truth, envy, and truthful market clearing bundle pricing.
\newblock In {\em Proceedings of the 7th Workshop on Internet and Network
  Economics}, WINE '11, pages 97--108, 2011.

\bibitem{Feldman2013}
M.~{Feldman}, H.~{Fu}, N.~{Gravin}, and B.~{Lucier}.
\newblock Simultaneous auctions are (almost) efficient.
\newblock In {\em STOC}, 2013.

\bibitem{Gale2001}
Ian Gale and Mark Stegeman.
\newblock {Sequential Auctions of Endogenously Valued Objects}.
\newblock {\em Games and Economic Behavior}, 36(1):74--103, July 2001.

\bibitem{Gul1999}
Faruk Gul and Ennio Stacchetti.
\newblock Walrasian equilibrium with gross substitutes.
\newblock {\em Journal of Economic Theory}, 87(1):95 -- 124, 1999.

\bibitem{Hassidim2011}
A.~Hassidim, Haim Kaplan, Yishay Mansour, and Noam Nisan.
\newblock Non-price equilibria in markets of discrete goods.
\newblock In {\em EC'11}.

\bibitem{huang12}
Zhiyi Huang, Nikhil~R. Devanur, and David~L. Malec.
\newblock Sequential auctions of identical items with budget-constrained
  bidders.
\newblock {\em CoRR}, abs/1209.1698, 2012.

\bibitem{Lehmann2001}
Benny Lehmann, Daniel Lehmann, and Noam Nisan.
\newblock Combinatorial auctions with decreasing marginal utilities.
\newblock In {\em EC}, 2001.

\bibitem{McAfee1993}
R.~Preston McAfee.
\newblock Mechanism design by competing sellers.
\newblock {\em Econometrica}, 61(6):pp. 1281--1312, 1993.

\bibitem{Milgrom1982a}
P.R. Milgrom and R.J. Weber.
\newblock {A theory of auctions and competitive bidding II}, 1982.

\bibitem{DiningBidder}
Renato Paes~Leme, Vasilis Syrgkanis, and \'{E}va Tardos.
\newblock The dining bidder problem: a la russe et a la francaise.
\newblock SIGecom Exchanges, Vol 11-2, 2012.

\bibitem{PaesLeme2012}
Renato Paes~Leme, Vasilis Syrgkanis, and \'{E}va Tardos.
\newblock Sequential auctions and externalities.
\newblock In {\em SODA}, 2012.

\bibitem{Rodriguez2009}
G.E. Rodriguez.
\newblock {Sequential Auctions with Multi-Unit Demands}.
\newblock {\em Theoretical Economics}, 9(1), 2009.

\bibitem{Syrgkanis2012a}
Vasilis Syrgkanis and Eva Tardos.
\newblock Bayesian sequential auctions.
\newblock In {\em EC}, 2012.

\bibitem{Syrgkanis2013}
Vasilis Syrgkanis and Eva Tardos.
\newblock Composable and efficient mechanisms.
\newblock In {\em STOC}, 2013.

\bibitem{Weber2000}
R.J. Weber.
\newblock {Multiple-object auctions}.
\newblock {\em Discussion Paper 496, Kellog Graduate School of Management,
  Northwestern University}, 1981.

\end{thebibliography}

\appendix

\section{Iterated Elimination of Weakly Dominated Strategies}
\label{app:iewds}

When considering subgame perfect equilibria of sequential item auctions, we wish to restrict our attention to ``natural'' equilibria, that exclude (for example) dominated overbidding strategies.  We therefore consider a natural and well-studied refinement of the set of subgame perfect equilibria: those that survive iterated elimination of weakly dominated strategies (IEWDS).  A strategy $s$ is \emph{weakly dominated} by a strategy $s'$ if, for every profile of other players' strategies $s_{-i}$, we have $u_i(s,s_{-i}) \leq u_i(s', s_{-i})$, and moreover there exists some $s_{-i}$ such that $u_i(s, s_{-i}) < u_i(s', s_{-i})$.  We can now define what it means for a strategy profile to survive iterated elimination of weakly dominated strategies.

\begin{definition}
Given an $n$-player game defined by strategy sets $S_1, \dotsc, S_n$ and utilities $u_i \colon S_1 \times \dotsc \times S_n \to \mathbb{R}$ we define a \emph{valid procedure for eliminating weakly dominated strategies} as a sequence $\{S_i^t\}$ such that for each $t$ there is an $i$ such that $S_j^t = S_j^{t-1}$ for $j \neq i$, $S_i^t \subseteq S_t^{t-1}$, and for all $s_i \in S_i^{t-1} \backslash S_i^t$ there is some $s_i' \in S_i^t$ such that $u_i(s_i', s_{-i}) \geq u_i(s_i, s_{-i})$ for all $s_{-i} \in \prod_{j \neq i}S_j^t$ and the inequality is strict for at least one $s_{-i}$.  We say that a strategy profile $s$ survives iterated elimination of weakly dominated strategies (IEWDS) if, for any valid procedure $\{S_i^t\}$, $s_i \in \cap_t S_i^t$.
\end{definition}

%We will focus on subgame perfect equilibria of sequential first-price item auctions that survive IEWDS.  It is shown in \cite{PaesLeme2012} that there always exists such an equilibrium. 
%%A characterization of such equilibria is also given in \cite{PaesLeme2012}.  
%We note one necessary property of an equilibrium satisfying IEWDS: in every subgame beginning at a time $t = m$ (i.e., when the last item is being sold), for every possible bidding history up to that round, each player will bid no more than his marginal value for the final item.  In other words, no player can credibly threaten to overbid on the last item for sale.

\section{Augmenting Path Orderings: An Example}
\label{app:conj}

In \cite{PaesLeme2012}, it was shown that there exist single-valued unit-demand auctions in which inefficient outcomes can occur when items are sold sequentially in an arbitrary order.  In this section we motivate that augmenting path ordering by showing that, for this example, the efficient outcome occurs when the items are sold according to their augmenting path order.

We begin by recalling the example.  There are three items, $\{A,B,C\}$, and 4 players $\{a,b,c,d\}$.  We fix an arbitrarily small constant $\epsilon > 0$.  Recall that the valuation of each player is specified by a real value $v$ and a set $S$ of items of interest; the player then has value $v$ for any item in $S$ and value $0$ for any other item.  The valuations in our example are given by:
\begin{itemize}
\item $v_a = \epsilon$ and $S_a = \{A\}$, 
\item $v_b = 1$ and $S_b = \{A, C\}$, 
\item $v_c = 1$ and $S_c = \{B, C\}$, and
\item $v_d = 1-\epsilon$ and $S_d = \{B\}$.
\end{itemize}
The welfare-optimal allocation is $(x_a, x_b, x_c, x_d) = (\emptyset, \{A\}, \{C\}, \{B\})$, for a social welfare of $3-\epsilon$.  The VCG prices are $(p_a, p_b, p_c, p_d) = (0, \epsilon, \epsilon, \epsilon)$.  Note that, in the terminology of Section \ref{sec:ordering}, player $a$ is the price-setter for each of the other players.  In \cite{PaesLeme2012} it is shown that if the items are auctioned in the order $(A,B,C)$, then the unique subgame perfect equilibrium that survives IEWDS leads to an inefficient outcome.

What are the augmenting path orderings in this example?  In this example, the augmenting path graph is a line, given by nodes $(a,A,b,C,c,B,d)$ in that sequence.  There is therefore a unique augmenting path ordering over the items: the order $(B, C, A)$.

We can now solve for the subgame perfect equilibrium of the auction when items are sold in this order.  We do so by analyzing the item auctions in reverse order.  When item $A$ is sold, the outcome depends on whether or not player $b$ won item $C$: if so, player $a$ will win item $A$ for a price of $0$, yielding $u_a = \epsilon$; if not, then player $b$ will win item $A$ for a price of $\epsilon$, yielding $u_b = 1 - \epsilon$ and $u_a = 0$.  This allows us to determine the outcome of the auction for item $C$: because player $b$ knows that she can win item $A$ for a price of $\epsilon$, she is willing to bid no more than $\epsilon$ on item $C$.  Thus, if player $c$ did not previously win item $B$, then player $c$ can win item $C$ with a bid of $\epsilon^+$, yielding $u_c = 1 - \epsilon$.  This ultimately allows us to determine the outcome of the first auction, the auction for item $B$.  Because player $c$ knows that he can win item $B$ for a price of $\epsilon$, she is willing to bid no more than $\epsilon$ on item $B$.  Since player $d$ obtains positive utility only if she wins item $B$, she is willing to bid as much as $v_d = 1 - \epsilon$ on item $B$.  We therefore have that player $d$ will choose to win item $B$ with a bid of $\epsilon^+$, obtaining utility $u_d = 1 - 2\epsilon$.  Applying our analysis of the subsequent auctions, we conclude that bidder $c$ will win item $C$ for a price of $\epsilon$, and then bidder $b$ will win $A$ for a price of $\epsilon$.

Note that this subgame perfect equilibrium, which is the unique equilibrium in undominated strategies, precisely implements the VCG outcome.  Moreover, our analysis extends easily to other values of $v_b, v_c$, and $v_d$, as long as they are all greater than $\epsilon$.

\section{Not all Augmenting Path Orderings lead to Efficiency}
\label{app:cex}

We now show that if the augmenting path graph is not a line, then some augmenting path orderings do not result in an efficient outcome, even if valuations are unit demand single-valued.

The example is as follows.  There are $3$ items, say $\{A, B, C\}$.  There are $4$ players.  Player $1$ wants all items and has value $1$.  Player $2$ wants only item $B$ and has value $2$.  Player $3$ wants item $B$ or $C$ and has value $3$.  Player $4$ want item $A$ or $C$ and has value $4$.

In this example, the VCG outcome is $(x_1, x_2, x_3, x_4) = (\emptyset, B, C, A)$, and the VCG prices are $(p_1, p_2, p_3, p_4) = (0, 1, 1, 1)$.  The augmenting path graph is a tree with player $1$ at the root, each item a child of player $1$, and each remaining player $i$ being the child of item $x_i$.  For this graph, every order over the items is an augmenting path order.

Suppose the items are sold in the order $(A, B, C)$.  In the VCG outcome, player $4$ obtains utility $v_4 - p_4 = 3$.  In the sequential play corresponding to the VCG outcome, players $1$ and $4$ both bid their values on item $A$.  Consider the following deviation by player $4$.  When item $A$ is sold, he bids $0$, causing player $1$ to win item $A$.  Item $B$ will sell next; players $2$ and $3$ will bid on it.  Consider what would happen if player $2$ wins item $B$: in this case, players $3$ and $4$ both bid their values on item $C$, and hence player $4$ wins $C$ and player $3$ ends with utility $0$.  We conclude that player $3$ prefers to win item $B$ at any price less than $v_3 = 3$, and hence will bid $3$ on item $B$, winning it.  Thus, when item $C$ is sold, only player $4$ places a non-zero bid, winning the item at price $0$.  We conclude that, after this deviation, player $4$ obtains utility $4$, and therefore the VCG outcome is not a SPE.

%\blcomment{While writing this up I noticed the following.  There is some flexibility in how we build the augmenting path graph.  (See the "tree construction" point in the previous section).  To get the graph described above, we need to build the children of player $1$ in a particular order: item $B$ first, then item $C$, then item $A$.  Under any other ordering, we get a different tree.  For example, if we place item $A$ first, then we get a line graph; if we place item $C$ first, we get $B$ being a descendent of $C$.  The deviation described above fails to hold for those graphs.}
%
%\blcomment{So here's a thought: the augmenting path order we choose will always order subtrees in the same order that the children were added to construct the tree.  It seems as though, under this ordering of the items, a player $i$ will only ever desire items that are either ancestors of $i$ in the tree, or else have already sold when it's time to sell item $x_i$.  Is this true?  If so, it seems like the proof sketch for the ``line graph'' case should work.}

\end{document}